\documentstyle{article}
\newcommand{\EcoLab}{{\sffamily\slshape
    \mbox{\raisebox{.5ex}{Eco}\hspace{-.4em}{\makebox[.5em]{L}ab}}}}

\title{\EcoLab: Agent Based Modeling for C++ programmers}
\author{Russell K. Standish and Richard Leow\\
High Performance Computing Support Unit\\
University of New South Wales, Sydney 2052, Australia\\
\{R.Standish,R.Leow\}@unsw.edu.au\\
http://parallel.hpc.unsw.edu.au/ecolab
}

\begin{document}
\maketitle

\begin{abstract}
\EcoLab{} is an agent based modeling system for C++ programmers,
strongly influenced by the design of Swarm. This paper is just a brief
outline of \EcoLab's features, more details can be found in other
published articles, documentation and source code from the \EcoLab{} website.
\end{abstract}

\section{\protect\EcoLab}
\EcoLab{} is an ABM system for C++ programmers. This is not the time
or place to debate the merits of C++ over any other object oriented
language. If you have chosen C++ as an implementation language for
your models because of performance, expressibility, familiarity or
compatibility with other software libraries, then ABM environments
such as Swarm or Repast offer little support to you. In this case, you
should consider \EcoLab.

\section{Scripting}
\EcoLab{} uses the Classdesc\cite{Madina-Standish01} object descriptor
technology. This provides a form of object reflection, or the ability
to query an object's internal structure at runtime. This may seem
obvious to a Java or Objective C programmer, as object reflection is
built into the language.

How is Classdesc used in \EcoLab{}? The user codes their entire model
as a class. Usually, there will only be one instantiated object of
that class (the model). Most model instance variables, and model
methods adhering to particular calling conventions are exposed to a
TCL interpreter. This allows great flexibility to configure different
sorts of experiments at runtime. For example, if your model class is:
\begin{verbatim}
class model_t {
public:
int tstep; double foo;
void step();
double average_something();
} model;
\end{verbatim}
then by inserting the macro call \verb+make_model(model)+ into your
code, the following TCL script is possible:
\begin{verbatim}
model.tstep 0 
model.foo 0
while {[model.tstep]<100000} {
   model.step
   if {[model.tstep]%1000==0} {puts stdout [model.average_something]}
}
\end{verbatim}
This initialises the instance variables, and runs the model for 100000
steps, writing out the average of something every 1000th
timestep. 

\section{GUI mode for exploration}

The TCL interpreter also has a complete GUI toolkit (Tk),
and a visualisation and analysis toolkit (BLT). It is possible to turn
the above script into a continuously updated plot by changing one
line:
\begin{verbatim}
   if {[model.tstep]%1000==0} {
      set av_something [model.average_something]
      plot av av_something
}
\end{verbatim}

With this ability to script experiments at runtime, one can use one
script for exploratory visualisation or debugging, and another for
batch production work. Items in common, such as the model's parameters
can be stored in a third file and included using TCL's \verb+source+
command.

The main point is that it is not necessary to be proficient at TCL
programming to be productive with \EcoLab. Many example scripts exist
in the source code that can be adapted for you use. However, a
proficient TCL programmer can exploit a large amount of functionality
to create some great visualisations --- using Tk's ability to handle
pixmaps for instance. The jellyfish simulation example provided as
part of the package is a case in point.

\section{Object Browser}
Using Classdesc to expose your C++ objects to the TCL environment has
some other interesting features. \EcoLab{} comes with an {\em object
  browser}, which allows the user to drill down into the model to see
why a particular object is doing what it is doing. The object browser
is \EcoLab's answer to Swarms probes. To use the object browser, just
click on the ``Object Browser'' button on the GUI toolbar. This will
pop a list of TCL procedures, and top level objects (procedures with ``.'' in
their name). Clicking on objects pops up another box containing the
procedures within that (corresponding to instance variables and
methods) and objects (compound instance variables). Clicking on a
procedure will execute that procedure (using any arguments you have
provided) and displays the result. You can also arrange to have the
procedure executed automatically every second to give a continuous
update of (say) and instance variable.

\section{Checkpoint/Restart}
By invoking Classdesc on your model in this way, TCL commands are
created for checkpointing and restarting your model, with no further
coding required of your model (provided the complete state of your
model is stored in the model object of course). So, if you are using a
high performance computing bureau, which forces you to continuously
checkpoint and restart your jobs to allow other peoples jobs to run,
you can write something like the following TCL script:
\begin{verbatim}
if [file exists checkpoint] {
   model.restart checkpoint
} else {
   source model-parms
}

while {[model.tstep]<100000} {
   model.step
   if {[cputime] > 10000} {
      model.checkpoint checkpoint
      exit
   }
}

if {[model.tstep]<100000} {
  exec qsub myjob.tcl
}
\end{verbatim}
which is self-submitting batch script in which each job only runs for
less that 3 hours, but the chain of jobs continues until the
calculation is complete.

\section{Parallel Processing}

\EcoLab{} provides support for parallel processing using the
MPI\cite{mpiref} distributed memory programming model. By turning on
the MPI flag at compile time, \EcoLab{} will start an interpreter on
each processor. Normally, your script will run on processor 0, but any
TCL command can be run on all processors simultaneously with the
\verb+parallel+ command. Plus, a method can be declared parallel, so
when invoked, it will be invoked on all processors simultaneously.
The {\em ClassdescMP}\cite{Standish-Madina03a} can be used to easily
code communication between processors, or the full power of MPI used.
At this point in time, it is the model implementor's responsibility to
arrange for objects to be distributed across the processors, however
an experimental project called {\em Graphcode}\cite{Madina02} will
allow arbitrary agents on an arbitrary grid topology (graph) to be
automatically distributed across parallel processors, even having
dynamic updating of the object distribution to optimise load balancing.

\section{Supported Machines}

\EcoLab{} is an open source project which depends on ANSI standard
C++, TCL, Tk and BLT. The primary development environment is Linux,
but by its adherence to standards, \EcoLab{} has been successfully
ported to Irix, Solaris, AIX, Tru64, Mac OSX and Windows (under
Cygwin), often using the native C++ compiler rather than gcc. It
should be noted that the Mac OSX port of \EcoLab{} was completed about
6 months ago, and took about 2 days to do, which is one area in which
it out competed Swarm. At present the interface depends on X-windows,
work on a native Aqua interface is being planned in the next 6 months.

\EcoLab{} has now been deployed for 5 distinctly different modelling
projects. Each project has its own requirements, which are fed back
into the core system where it is useful.

There are no plans for a Java interface. The aim of this project is to
provide an environment for C++ programmers, not Java programmers who
are already catered for with other packages.

We welcome contributers to the \EcoLab{} project. The source code is
managed by Peter Miller's AEGIS system, so you will need to email
Russell Standish to obtain a user login to access AEGIS. Anonymous readonly
access to the repository is however already available via the
\EcoLab{} website, where you can pick a version of  \EcoLab{} at any
revision level as a gzipped tarball. Unlike CVS, AEGIS ensures that
the code compiles, and that some consistency checks are performed at
code check in time. Plus whenever a branch is closed (\EcoLab.4.Dx as
opposed to \EcoLab.4.x.Dy), further tests are performed to ensure that
the code compiles and runs a test suite on all supported platforms.

\section*{Acknowledgments}
\EcoLab{} is supported by a generous grant from the {\em Australian
  Partnership for Advanced Computing} (APAC). Computing resources
  have been generously provided by
  {\em Australian Centre for Advanced Computing and Communications}
  (ac3) for supporting different operating systems and testing
  parallel programming features.


\begin{thebibliography}{1}

\bibitem{Madina02}
Duraid Madina.
\newblock Topology and complexity: From automata to agents.
\newblock In Namatame et~al., editors, {\em {Complex Systems} '02}, pages
  141--147. Chuo University, September 2002.
\newblock also to appear in {\em Complexity International}.

\bibitem{Madina-Standish01}
Duraid Madina and Russell~K. Standish.
\newblock A system for reflection in {C++}.
\newblock In {\em Proceedings of AUUG2001: Always on and Everywhere}.
  Australian Unix Users Group, 2001.

\bibitem{mpiref}
Marc Snir et~al.
\newblock {\em MPI: the complete reference}.
\newblock MIT Press, Cambridge, MA, 1996.

\bibitem{Standish-Madina03a}
R.K. Standish and D.~Madina.
\newblock {ClassdescMP}: Easy {MPI} programming in {C++}.
\newblock In {\em Proceedings of Intennational Conference on Computational
  Science 2003}, Lecture Notes in Computer Science, Berlin, 2003. Springer.
\newblock (accepted).

\end{thebibliography}

\end{document}